\documentstyle[aps,preprint,epsfig]{revtex}
\input epsf.tex
\def\DESepsf(#1 width #2){\epsfxsize=#2 \epsfbox{#1}}
\begin{document}
%
% SET UP DEFINITIONS IN HERE.  Make this one OFFICIAL MBC version!
%
%    6/25/97  all cplear stuff moved to cpleardef.tex
%
%     Latex stuff first:
%
% Set up better \caption command.  Both params are now mandatory
%\newcommand{\capt}[2]{ \begin{minipage}{0.8\textwidth} \caption[#1]{
%    \small #2} \end{minipage}  } 
%\newcommand{\capt}[3]{ \begin{minipage}{0.8\textwidth} \caption[#1]{
%    \small #2} \label{#3} \end{minipage}  } 
\newcommand{\ifm}[1]{\relax\ifmmode #1\else $#1$\fi}
\newcommand{\capt}[3]{ \begin{minipage}{0.8\textwidth} \caption[#1]{
      \renewcommand{\baselinestretch}{1} \small #2} \label{#3} \end{minipage}  } 
\newcommand{\beq}{\begin{equation}}
\newcommand{\eeq}{\end{equation}}
\newcommand{\beqn}{\begin{eqnarray}}
\newcommand{\eeqn}{\end{eqnarray}}
\newcommand{\bi}{\begin{itemize}}
\newcommand{\ei}{\end{itemize}}
\newcommand{\bd}{\begin{description}}
\newcommand{\ed}{\end{description}}
\newcommand{\bHuge}{\begin{Huge}}
\newcommand{\bhuge}{\begin{huge}}
\newcommand{\bLARGE}{\begin{LARGE}}
\newcommand{\bLarge}{\begin{Large}}
\newcommand{\blarge}{\begin{large}}
\newcommand{\eHuge}{\end{Huge}}
\newcommand{\ehuge}{\end{huge}}
\newcommand{\eLARGE}{\end{LARGE}}
\newcommand{\eLarge}{\end{Large}}
\newcommand{\elarge}{\end{large}}
\def \mc {\multicolumn}
\def \mspace {\mbox{ }\mbox{ }\mbox{ }\mbox{ }}
\newcommand{\degr}{\mbox{$^{\circ}$}}
\newcommand{\gt}{\ifm{>}}
\newcommand{\lt}{\ifm{<}}
\def \gtsim    {\relax\ifmmode{\mathrel{\mathpalette\oversim >}}
                  \else{$\mathrel{\mathpalette\oversim >}$}\fi}
\def \ltsim    {\relax\ifmmode{\mathrel{\mathpalette\oversim <}}
                  \else{$\mathrel{\mathpalette\oversim <}$}\fi}
\def\oversim#1#2{\lower4pt\vbox{\baselineskip0pt \lineskip1.5pt
            \ialign{$\mathsurround=0pt#1\hfil##\hfil$\crcr#2\crcr\sim\crcr}}}
\def \dk       {\relax\ifmmode{\rightarrow}\else{$\rightarrow$}4\fi}
\def \rpvdk    {\relax\ifmmode{\Rightarrow}\else{$\Rightarrow$}\fi}
\newcounter{hours}\newcounter{minutes}
\newcommand{\printtime}{%
 	\setcounter{minutes}{\time}%
        \theminutes  min (since midnight)}
%
%%%%%%%%%%%%%%%%%%%%%%%%%%%%%%%%%%%%%%%%%%%%%%%%%%%%%%%%%%%%%%%%%
%%%%% Sunil's definitions %%%%
%%%%%%%%%%%%%%%%%%%%%%%%%%%%%%%%%%%%%%%%%%%%%%%%%%%%%%%%%%%%%%%%%
%
\newcommand\newcolour[2]{\newenvironment{#1}%
    {
\ignorespaces}%
    {
}}
%\newcolour{teal}          {0.00 0.50 0.70}
%\newcolour{darkbluegreen} {0.00 0.40 0.40}
\newcolour{gblue}         {0.00 0.50 1.00}
%\newcolour{bluegreen}     {0.00 0.60 0.60}
%\newcolour{jade}          {0.25 0.50 0.40}
%\newcolour{magent}        {0.55 0.00 0.50}
%\newcolour{huntergreen}   {0.00 0.50 0.00}
%\newcolour{desertred}     {0.90 0.35 0.00}
%\newcolour{coltitle}      {0.00 0.40 0.50}
%\newcolour{colsection}    {0.00 0.40 0.50}
%
%
\newcommand{\black}{\color{black}}
\newcommand{\blue}{\color{blue}}
\newcommand{\cyan}{\color{cyan}}
\newcommand{\green}{\color{green}}
\newcommand{\magenta}{\color{magenta}}
\newcommand{\red}{\color{red}}
\newcommand{\vgap}{\vspace{0.5in}}
\newcommand{\hgap}{\hspace{0.5in}}
%
%
%%%%%   D0    %%%%%
%
\newcommand{\dzero}{\mbox{D\O}}
\newcommand{\Dzero}{\mbox{D\O}}
\newcommand{\DZERO}{\Dzero\ Collaboration}
%
%%%%%   CDF   %%%%%
%
\newcommand{\CDF}{CDF Collaboration}
\def \svx {SVX II}
\def \svxp {SVX$^{\prime}$}
%
%%%%% HEP/Paricles %%%%%
%
\newcommand{\qbar}   {\mbox{$\overline{q}$}}
\newcommand{\cbar}   {\mbox{$\overline{c}$}}
\newcommand{\bbar}   {\mbox{$\overline{b}$}}
\newcommand{\tbar}   {\mbox{$\overline{t}$}}
\newcommand{\ppbar}{\mbox{$p\overline{p}$}}
\newcommand{\ppb}{\mbox{$p\overline{p}$}}
\newcommand{\bbbar}{\mbox{$b\overline{b}$}}
\newcommand{\bbb}{\mbox{$b\overline{b}$}}
\newcommand{\ccbar}{\mbox{$c\overline{c}$}}
\newcommand{\ccb}{\mbox{$c\overline{c}$}}
\newcommand{\qqbar}{\mbox{$q\overline{q}$}}
\newcommand{\ttbar}{\mbox{$t\overline{t}$}}
\newcommand{\ttb}{\mbox{$t\overline{t}$}}
\newcommand{\wpm}{\mbox{$W^{\pm}$}}
\newcommand{\Wpm}{\mbox{$W^{\pm}$}}
\newcommand{\Wp}{\mbox{$W^{+}$}}
\newcommand{\Wm}{\mbox{$W^{-}$}}
\newcommand{\zzero}{\mbox{$Z^0$}}
\newcommand{\zz}{\mbox{$Z^0$}}
\newcommand{\Zz}{\mbox{$Z^0$}}
\newcommand{\ee}{\mbox{$ee$}}
\newcommand{\wb}{\ifm{W}}
\newcommand{\zb}{\ifm{Z}}
\newcommand{\lnu}{\mbox{$\ell \nu$}}
\newcommand{\emu}{\mbox{$e\mu$}}
\newcommand{\mumu}{\mbox{$\mu\mu$}}
\newcommand{\mets}{\mbox{${E\!\!\!\!/_T}$}}
\newcommand{\met}{\mbox{${E\!\!\!\!/_T}$}}
\newcommand{\MET}{\mbox{${E\!\!\!\!/_T}$}}	% T.Kamon (June 1997)
\newcommand{\mpt}{\mbox{${p\!\!\!/_T}$}}
\newcommand{\rp}{\mbox{${R_p}$}}
\newcommand{\rpc}{\mbox{${R_p}$}}
\newcommand{\rpv}{\mbox{${R\!\!\!\!\!\:/_p}$}}
\newcommand{\lnv}{\mbox{${L\!\!\!\!\!\:/}$}}
\newcommand{\bnv}{\mbox{${B\!\!\!\!\!\:/}$}}
\newcommand{\lamp}{\mbox{${\lambda^{\prime}_{121}}$}}
\newcommand{\lum}{\mbox{${\cal L}$}}
\newcommand{\intlum}{\mbox{${ \int {\cal L} \; dt}$}}
\newcommand{\pt}{\mbox{$p_{T}$}}
\newcommand{\pT}{\mbox{$p_{T}$}}
\newcommand{\Pt}{\mbox{$p_{T}$}}
\newcommand{\ipt}{\mbox{$p^{-1}_{{\rm T}}$}}
\newcommand{\ptcut}{\mbox{$p_{{\rm T}}$-Cut}}
\newcommand{\et}{\mbox{$E_T$}}
\newcommand{\Et}{\mbox{$E_T$}}
\newcommand{\ET}{\mbox{$E_T$}}
\newcommand{\sigpt}{\mbox{$\sigma_{\pT}/\pT$}}
\newcommand{\dphi}{\mbox{$\Delta\varphi$}}
\newcommand{\Ht}{\ifm{H_T}}
\newcommand{\abseta}{\ifm{|\eta|}}
%
%
%%%%% HEP/Ractions %%%%%
%
\newcommand{\zee}{\mbox{$Z^{0} \rightarrow e^{+} e^{-}$}}
\newcommand{\zmumu}{\mbox{$Z^{0} \rightarrow \mu^{+} \mu^{-}$}}
%
%%%%  mSUGRA parameters %%%%%
%
\newcommand{\azero}{\ifm{A_0}}
\newcommand{\tanb}{\ifm{\tan\beta}}
\newcommand{\mzero}{\ifm{m_0}}
\newcommand{\mhalf}{\ifm{m_{1/2}}}
\newcommand{\mupos}{\ifm{\mu > 0}}
\newcommand{\muneg}{\ifm{\mu < 0}}
%
%%%%  SUSY  %%%%%
%
\newcommand{ \photino}{\mbox{$\tilde{\gamma}$}}
\newcommand{ \gravitino} {\mbox{$\tilde{G}$}}
\newcommand{ \gluino} {\mbox{$\tilde{g}$}}
\newcommand{ \squark} {\mbox{$\tilde{q}$}}
\newcommand{ \squarkb} {\mbox{$\bar{\tilde{q}}$}}
\newcommand{ \slepton} {\mbox{$\tilde{\ell}$}}
\newcommand{ \seleR} {\mbox{$\tilde{e}_{R}$}}
\newcommand{ \seleRpm} {\mbox{$\tilde{e}_{R}^{\pm}$}}
\newcommand{ \seleL} {\mbox{$\tilde{e}_{L}$}}
\newcommand{ \seleLpm} {\mbox{$\tilde{e}_{L}^{\pm}$}}
\newcommand{ \sele} {\mbox{$\tilde{e}$}}
\newcommand{ \selepm} {\mbox{$\tilde{e}^{\pm}$}}
\newcommand{ \selep} {\mbox{$\tilde{e}^{+}$}}
\newcommand{ \selem} {\mbox{$\tilde{e}^{-}$}}
\newcommand{ \smu} {\mbox{$\tilde{\mu}$}}
\newcommand{ \smuR} {\mbox{$\tilde{\mu}_R$}}
\newcommand{ \smuRpm} {\mbox{$\tilde{\mu}_R^{\pm}$}}
\newcommand{ \smupm} {\mbox{$\tilde{\mu}^{\pm}$}}
\newcommand{ \smup} {\mbox{$\tilde{\mu}^{+}$}}
\newcommand{ \smum} {\mbox{$\tilde{\mu}^{-}$}}
\newcommand{ \stauone} {\mbox{$\tilde{\tau}_{1}$}}
\newcommand{ \stauonep} {\mbox{$\tilde{\tau}_{1}^{+}$}}
\newcommand{ \stauonem} {\mbox{$\tilde{\tau}_{1}^{-}$}}
\newcommand{ \stauonepm} {\mbox{$\tilde{\tau}_{1}^{\pm}$}}
\newcommand{ \stau} {\mbox{$\tilde{\tau}$}}
\newcommand{ \stauL} {\mbox{$\tilde{\tau}_{L}$}}
\newcommand{ \stauR} {\mbox{$\tilde{\tau}_{R}$}}
\newcommand{ \staupm} {\mbox{$\tilde{\tau}^{\pm}$}}
\newcommand{ \staup} {\mbox{$\tilde{\tau}^{+}$}}
\newcommand{ \staum} {\mbox{$\tilde{\tau}^{-}$}}
\newcommand{ \snu}     {\mbox{$\tilde{\nu}$}}
\newcommand{ \sneutrino} {\mbox{$\tilde{\nu}$}}
\newcommand{ \usquark} {\mbox{$\tilde{u}$}}
\newcommand{ \dsquark} {\mbox{$\tilde{d}$}}
\newcommand{ \csquark} {\mbox{$\tilde{c}$}}
\newcommand{ \ssquark} {\mbox{$\tilde{s}$}}
\newcommand{ \scharm} {\mbox{$\tilde{c}$}}
\newcommand{ \csquarkl} {\mbox{$\tilde{c}_L$}}
\newcommand{ \csquarklb} {\mbox{$\bar{\tilde{c}}_L$}}
\newcommand{ \bcsquarkl} {\mbox{$\bar{\tilde{c}}_L$}}
\newcommand{ \bsquark} {\mbox{$\tilde{b}$}}
\newcommand{ \tsquark} {\mbox{$\tilde{t}$}}
\newcommand{ \sbottom}    {\mbox{$\tilde{b}$}}
\newcommand{ \sbottomb}    {\mbox{$\bar{\tilde{b}}$}}
\newcommand{ \sbottomone}    {\mbox{$\tilde{b}_{1}$}}
\newcommand{ \sbottomoneb}    {\mbox{$\bar{\tilde{b}}_{1}$}}
\newcommand{ \sstop}    {\mbox{$\tilde{t}$}}
\newcommand{ \stopo}   {\mbox{$\tilde{t}_1$}}
\newcommand{ \stopone}    {\mbox{$\tilde{t}_{1}$}}
\newcommand{ \stoponeb}    {\mbox{$\bar{\tilde{t}}_{1}$}}
\newcommand{ \stoptwo}    {\mbox{$\tilde{t}_{2}$}}
\newcommand{ \stoptwob}    {\mbox{$\bar{\tilde{t}}_{2}$}}
\newcommand{\sfermion}{\mbox{$\widetilde f$}}
\newcommand{ \cscsb}   {\mbox{$\csquarkl\bcsquarkl$}}
\newcommand{ \sqsq}   {\mbox{$\squark\squark$}}
\newcommand{ \sqsqb}   {\mbox{$\squark\squarkb$}}
\newcommand{ \ssb}     {\mbox{$\squark\overline{\squark}$}}
\newcommand{ \ststbone}   {\mbox{$\stopone\stoponeb$}}
\newcommand{ \ttbone}   {\mbox{$\stopone\stoponeb$}}
\newcommand{ \ttbtwo}   {\mbox{$\stoptwo\stoptwob$}}
\newcommand{ \cgino}{\mbox{$\tilde{\chi}^{\pm}$}}
\newcommand{ \nlino}{\mbox{$\tilde{\chi}^{0}$}}
\newcommand{ \ngino}{\mbox{$\tilde{\chi}^{0}$}}
\newcommand{ \wino}   {\mbox{$\tilde{W}^{\pm}$}}
\newcommand{ \zino}   {\mbox{$\tilde{Z}$}}
\newcommand{ \schi }{\mbox{$\tilde{\chi}$}}
\newcommand{ \lsp}    {\mbox{$\tilde{\chi}_{1}^{0}$}}
\newcommand{ \schionezero }{\mbox{$\tilde{\chi}_{1}^{0}$}}
\newcommand{ \schitwozero }{\mbox{$\tilde{\chi}_{2}^{0}$}}
\newcommand{ \schithreezero }{\mbox{$\tilde{\chi}_{3}^{0}$}}
\newcommand{ \schifourzero }{\mbox{$\tilde{\chi}_{4}^{0}$}}
\newcommand{ \schionepm }{\mbox{$\tilde{\chi}_{1}^{\pm}$}}
\newcommand{ \schionep }{\mbox{$\tilde{\chi}_{1}^{+}$}}
\newcommand{ \schionem }{\mbox{$\tilde{\chi}_{1}^{-}$}}
\newcommand{ \schitwopm }{\mbox{$\tilde{\chi}_{2}^{\pm}$}}
\newcommand{ \schitwop }{\mbox{$\tilde{\chi}_{2}^{+}$}}
\newcommand{ \schitwom }{\mbox{$\tilde{\chi}_{2}^{-}$}}
\newcommand{ \chizero}{\mbox{$\tilde{\chi}_{1}^0$}}
\newcommand{ \chione }{\mbox{$\tilde{\chi}_{1}^{\pm}$}}
\newcommand{ \chionemp }{\mbox{$\tilde{\chi}_{1}^{\mp}$}}
\newcommand{ \chipm  }{\mbox{$\tilde{\chi}^{\pm}$}}
\newcommand{ \chimp  }{\mbox{$\tilde{\chi}^{\mp}$}}
\newcommand{ \chip  }{\mbox{$\tilde{\chi}^{+}$}}
\newcommand{ \chim  }{\mbox{$\tilde{\chi}^{-}$}}
\newcommand{ \chionep   }{\mbox{$\tilde{\chi}_{1}^{+}$}}
\newcommand{ \chionem   }{\mbox{$\tilde{\chi}_{1}^{-}$}}
\newcommand{ \chitwo }{\mbox{$\tilde{\chi}_{2}^0$}}
\newcommand{ \chichi} {\mbox{$\chione\chitwo$}}
\newcommand{ \mgluino} {\mbox{$M(\gluino)$}}
\newcommand{ \msquark} {\mbox{$M(\squark)$}}
\newcommand{ \mcsl} {\mbox{$M(\csquarkl)$}}
\newcommand{ \msbotone} {\mbox{$M(\tilde{b}_{1})$}}
\newcommand{ \mstopone} {\mbox{$M(\tilde{t}_{1})$}}
\newcommand{ \mstopo}   {\mbox{$M(\tilde{t}_1)$}}
\newcommand{ \sqmass }{\mbox{$M_{\tilde{q}}$}}
\newcommand{ \glmass }{\mbox{$M_{\tilde{g}}$}}
\newcommand{ \hmass  }{\mbox{$M_{{H}^{\pm}}$}}
\newcommand{ \mchio}{\mbox{$M(\chizero)$}}
\newcommand{ \mchione}{\mbox{$M(\schionepm)$}}
%
%%%%%%  OTHER    %%%%%%%
%
\newcommand{\dagg}[1]{ #1 ^{\dagger}}
% Define modulus command to give
% two vertical bars.
% Must be used in math mode.
\newcommand{\modulus}[1]{\left| #1 \right|}
\newcommand{\paren}[1]{\left( #1 \right)}
\newcommand{\ave}[1]{\left\langle #1 \right\rangle}
\newcommand{\mods}[1]{\left| #1 \right|^2}
\newcommand{\bra}[1]{\langle #1 |}
\newcommand{\ket}[1]{| #1 \rangle}
\newcommand{\braket}[2]{\langle #1 | #2 \rangle}
\newcommand{\bracket}[3]{\langle #1 | #2 | #3 \rangle}
\newcommand{\sqbs}[1]{\left[ #1 \right] }
\newcommand{\rea}[1]{\mbox{$\,{\rm Re}\left( #1 \right)$}}
\newcommand{\ima}[1]{\mbox{$\,{\rm Im}\left( #1 \right)$}}
\def \sp       {\relax\ifmmode{\;}\else{$\;$}47\fi}	% T.Kamon (June 1997)
%
%  standard model PARTICLES
%
\newcommand{\pelp}{\mbox{$e^+$}}
\newcommand{\pelm}{\mbox{$e^-$}}
\newcommand{\pelpm}{\mbox{$\rm{e}^{\pm}$}}
\newcommand{\plp}{\mbox{$l^+$}}
\newcommand{\plm}{\mbox{$l^-$}}
\newcommand{\pmup}{\mbox{$\mu^+$}}
\newcommand{\pmum}{\mbox{$\mu^-$}}
\newcommand{\pprp}{\mbox{$\rm{p}^+$}}
\newcommand{\pprm}{\mbox{$\rm{p}^-$}}
\newcommand{\pprb}{\mbox{${\overline{p}}$}}
\newcommand{ \lpm    }{\mbox{$\ell^{\pm}$}}
\newcommand{ \lmp    }{\mbox{$\ell^{\mp}$}}
\newcommand{ \lplm    }{\mbox{$\ell^{+} \ell^{-}$}}
\newcommand{ \epm    }{\mbox{$e^{\pm}$}}
\newcommand{ \emp    }{\mbox{$e^{\mp}$}}
\newcommand{ \mupm    }{\mbox{$\mu^{\pm}$}}
\newcommand{ \mump    }{\mbox{$\mu^{\mp}$}}
\newcommand{ \eplus    }{\mbox{$e^{+}$}}
\newcommand{ \eminus    }{\mbox{$e^{-}$}}
\newcommand{ \lplus  }{\mbox{$\ell^+$}}
\newcommand{ \lminus }{\mbox{$\ell^-$}}
%
%  Invariant masses, etc
%
\newcommand{\mll}{\mbox{$M(\ell\ell)$}}
\newcommand{\mjj}{\mbox{$M(jj)$}}
\newcommand{\mlljj}{\mbox{$M(\ell\ell jj)$}}
\newcommand{\dphill}{\mbox{$\Delta \phi (\ell\ell)$}}
\newcommand{\dphijj}{\mbox{$\Delta \phi (jj)$}}
%
%       Units
%
\newcommand{\mev}  {\mbox{${\rm MeV}/c$}}
\newcommand{\mevc} {\mbox{${\rm MeV}/c^2$}}
\newcommand{\gev}  {\mbox{${\rm GeV}$}}
\newcommand{\gevc} {\mbox{${\rm GeV}/c$}}
\newcommand{\pgev} {\mbox{${\rm GeV}$}}
\newcommand{\gevcc}{\mbox{${\rm GeV}/c^2$}}
\newcommand{\mmev}{\mbox{${\rm MeV}/c^2$}}
\newcommand{\mgev}{\mbox{${\rm GeV}$}}
\newcommand{\tev}  {\mbox{${\rm TeV}$}}
\newcommand{\mtev}{\mbox{${\rm TeV}$}}
\newcommand{\tevcc}{\mbox{${\rm TeV}/c^2$}}
\newcommand{\cmtwo}{\mbox{${\rm cm}^2$}}
%
%   luminosity
%
\newcommand{\invnb}{\mbox{${\rm nb}^{-1}$}}
\newcommand{\ipb}{\mbox{${\rm pb}^{-1}$}}
\newcommand{\invpb}{\mbox{${\rm pb}^{-1}$}}
\newcommand{\ifb}{\mbox{${\rm fb}^{-1}$}}
\newcommand{\invfb}{\mbox{${\rm fb}^{-1}$}}
\newcommand{\lumin}{\mbox{${\rm cm}^{-2}{\rm s}^{-1}$}}
%
%  Miscellaneous
%
\newcommand{\chis}{\mbox{$\chi^{2}$}}
\newcommand{\ie}{i.e.}
\newcommand{\eg}{e.g.}
\newcommand{\dedx}{\mbox{${\rm d}E/{\rm d}x$}}
\newcommand{\dndx}{\mbox{${\rm d}n/{\rm d}x$}}
\newcommand{\micron}{\mbox{$\mu{\rm m}$}}
\newcommand{\musec}{\mbox{$\mu {\rm s}$}}
\newcommand{\eps}{\mbox{$\epsilon$}}
\newcommand{\dxdy}{\mbox{${\rm d}X{\rm d}Y$}}
\newcommand{\sigrphi}{\mbox{$\sigma_{r \phi}$}}
\newcommand{\sigz}{\mbox{$\sigma_{z}$}}
\newcommand{\deltapp}{\mbox{$\Delta p/p$}} 
\newcommand{\epem}{\mbox{$e^+e^-$}}
\newcommand{\upum}{\mbox{$\mu^+\mu^-$}}
%
%
% Publishing papers
%
\def\Journal#1#2#3#4{{#1} {\bf #2}, #3 (#4)}
\def \PRL      {Phys. Rev. Lett.~}
\def \PR       {Phys. Rev.}
\def \PRD      {Phys. Rev. D}
\def \PL       {Phys. Lett.~}
\def \PLB      {Phys. Lett. B}
\def \ZPC      {Z. Phys. C}	% - Particles and Fields}
\def \NPB      {Nucl. Phys. B}
\def \NPD      {Nucl. Phys. D}
\def \PR       {Phys. Rep.~}
\def \INC      {Il Nuovo Cimento}
\def \NIM      {Nucl. Instrum. Methods}
\def \NIMA     {Nucl. Instrum. Methods Phys. Res. Sect. A}
\def \CPC	{Commput. Phys. Commun.}
\def \etal     {\relax\ifmmode{et \; al.}\else{$et \; al.$}\fi}
%
%
%------------------------------------------------------------------------------
\newcommand{\GEANT}{{\sc geant}}
\newcommand{\ISAJET}{{\sc isajet}}
\newcommand{\HERWIG}{{\sc herwig}}
\newcommand{\VECBOS}{{\sc vecbos}}
\newcommand{\SPYTHIA}{{\sc spythia}}
\newcommand{\PYTHIA}{{\sc pythia}}
\newcommand{\TAUOLA}{{\sc tauola}}
\newcommand{\QQ}{{\sc qq}}
\newcommand{\SHW}{{\sc shw}}
%------------------------------------------------------------------------------

\newlength{\pushupfigure}
\setlength{\pushupfigure}{-55.5pt}
\def \epsfin_v1#1#2{
	\vspace{\pushupfigure}
	\center
	\leavevmode
	\epsfxsize=#1
	\epsffile[20 143 575.75 698.75]{#2}
}
%---- 2/15/98 Noticed Latex does not recognize _ in the command. Therefore
% upgrading this command by epsfin_v2 does not work. It seems working in doing
% epsfA, epsfB, ... So let's indicate its version by a capital letter A, B, ...
%
% A = Same as \epsfin_v1
\def \epsfA#1#2{
	\vspace{\pushupfigure}
	\center
	\leavevmode
	\epsfxsize=#1
	\epsffile[20 143 575.75 698.75]{#2}
}

% B = Bounding box is the same as the plot box, not the box surrounding a plot
% Also \pushupfigure is removed.
\def \epsfB#1#2{
%	\vspace{\pushupfigure}
	\center
	\leavevmode
	\epsfxsize=#1
	\epsffile[75.5 198.5 520.25 643.25]{#2}
}

\def \bitm \begin{itemize}
\def \eitm \end{itemize}

%---- Reserved for the future upgrade
\def \epsfC#1#2{
%	\vspace{\pushupfigure}
%	\center
	\leavevmode
	\epsfxsize=#1
	\epsffile[75.5 198.5 520.25 643.25]{#2}
}

\def \epsfD#1#2{
%	\vspace{\pushupfigure}
	\center
	\leavevmode
	\epsfxsize=#1
	\epsffile{#2}
}

\def \epsfE#1#2{
%	\vspace{\pushupfigure}
	\leavevmode
	\epsfxsize=#1
	\epsffile{#2}
}

\def \mspace {\mbox{ }\mbox{ }\mbox{ }\mbox{ }\mbox{ }\mbox{ }}

\newlength{\figsize}
\setlength{\figsize}{\textwidth}

\draft

%@twocolumnfalse\endcsname
\preprint{\hbox{CTP-TAMU-35-00}}
\title{Prospect for   Searches for
Gluinos and Squarks
at a\\ Tevatron Tripler}
\author{ V. Krutelyov$^{1}$,
  R. Arnowitt$^{2}$,
  B. Dutta$^{2}$,\\
  T. Kamon$^{1}$,
  P. McIntyre$^{1}$,
  Y. Santoso$^{2}$}
\address{$^1$Department of Physics, Texas A$\&$M University, College Station, TX 
77843-4242,\\$^2$Center for Theoretical Physics, Department of Physics, Texas A$\&$M
University, \\ College Station, TX  77843-4242.}
\date{November, 2000}
\maketitle
\thispagestyle{empty}

\begin{abstract}  
We examine the  discovery potential for SUSY new physics at a $p{\bar
p}$
collider upgrade of Tevatron with $\sqrt s$ = 5.4 TeV and luminosity 
${\cal L}\simeq 4\times 10^{32}$ cm$^{-2}$s$^{-1}$ (the
Tripler). We consider the reach for gluinos $(\tilde g)$ and squarks 
($\tilde q$) using the
experimental signatures with large missing transverse energy (\met) of
jets~+~\met\ and
$1\ell$~+~jets~+~\met\ (where $\ell$=electron or muon) within the framework of minimal
supergravity. The Tripler's strongest reach for the  gluino is
 1060 GeV for the jets~+~\met\  channel and 1140 GeV for the 
$1\ell$~+~jets~+~\met\ channel for 30 fb$^{-1}$ of integrated luminosity (approximately
two years running time). This is to be compared with the Tevatron where the
reach is 440(460) GeV in the  jets~+~\met\ channel for 15(30) fb$^{-1}$
of integrated luminosity.
\end{abstract}

\vskip1.0in

\newpage
\section{Introduction}
\label{sec:intro}
The Tripler \cite{tripler} is a proposed energy upgrade of the Tevatron,
in which its ring of 4 Tesla NbTi superconducting magnets would be
replaced by a ring of 12 Tesla Nb$_{3}$Sn magnets.
Thanks to improvements in  Nb$_{3}$Sn technology and in dipole design
methodology, it is now possible to extend dipole fields up to
and beyond  12 Tesla.
Prototype magnets are being developed using several different methodologies
at Brookhaven National Lab \cite{BNLmagnet},
Fermilab \cite{FNALmagnet},
Lawrence Berkeley National Lab \cite{LBNLmagnet}, and
Texas A\&M University \cite{TAMUmagnet}.

The rationale for the Tripler is that the upgrade opens an energy window
in which the particles of the Higgs sector  and
new physics  are expected to be produced in a mass range of
$\ltsim\ 1\ \mtev$.
The Tripler furthermore accesses this energy window primarily through
quark-antiquark annihilation and gluon fusion, whereas the Large Hadron Collider (LHC)
will access a similar window primarily through gluon fusion and
gluon-quark/quark-quark interaction.
The proposed Next Linear Collider (NLC)
with its center-of-mass (c.m.) energy of 500 \gev\ \cite{bag}
will access a  limited energy window
through $e^+ e^-$ annihilation, but with more precise measurements
of the parameters of theories.
Complimentarity has often proved vital in understanding
new phenomena at the high-energy frontier.
The Tripler would use the existing tunnel, existing $\bar{p}$ source and
injector accelerators, and existing detectors CDF and D\O\ with
minimal changes. With a luminosity of about 
$4\times 10^{32}$ cm$^{-2}$s$^{-1}$ and live time of $2\times 10^{7}$ sec/year,
the Tripler would deliver about 8 fb$^{-1}$/year for each detector\cite{tripler}.

The present paper is concerned with evaluating the
reach of the Tripler for new physics.
Reference\cite{tripler_wisc} analyzed the signals
for the Standard Model (SM) Higgs boson at the Tripler and compared
 them with
those at the LHC  \cite{atlas_tdr,cms_higgs}.
It is remarkable that the Tripler can discover a Higgs boson
up to 680 (600) \mgev\ mass
with 40 (10) \invfb\ of integrated luminosity,
which is close to the triviality  upper bound
of 710 \mgev\ \cite{smhiggs_trivbound}.
A light  Higgs boson
($\ltsim\ 130\ \mgev$) would be 
accessible via $WWH$ coupling with 7.5 \invfb\ at the Tripler, while
its production at the LHC proceeds predominantly via processes involving
Yukawa couplings.

Another important benchmark for the physics is
the potential to discover the particles of  supersymmetry (SUSY) \cite{mssm}.
There have been  extensive analyses of the discovery potential
for SUSY particles, based on minimal spergravity (mSUGRA) model\cite{msugra} or minimal
Supersymmetric Standard Model (MSSM), at the Tevatron
\cite{tev2000_baer,tev2000_mrenna,tev2000,Baer_large_tanB,Barger_large_tanB,baer2000,run2shw_tau,Accomando,anderson,regina,run2shw}
and at the LHC \cite{atlas_tdr,cms_susy}. In the Tripler case,
SUSY studies on $\ppbar \to 3 \ell + \met\ + X$
(dominantly from  $\schionepm\schitwozero$ production)
and
$\ppbar \to  \ell^{\pm} \ell^{\pm}$ + jets + \met\
(dominantly from $\gluino\gluino / \gluino\squark$ production)
have been carried out in Ref. \cite{tripler_wisc}. In 
this paper we present a comparative study of the discovery reaches for
gluinos and squarks with
large missing transverse energy (\met). We examine the signals from 
jets + \met\ and $1\ell$ + jets + \met\ ($\ell$=electron or muon) 
at the Tripler and compare these signals at the Tevatron.
\section{mSUGRA Model}
\label{sec:susymodel}

To test the reach of the Tripler for gluinos ($\tilde g$) and squarks ($\tilde
q$), we consider SUSY models for which grand unification of the gauge coupling
constants occur at a GUT scale $M_G\equiv 2\times 10^{16}$ GeV. These models
are consistent with the LEP measurements of $\alpha_i$($i$=1,2,3) at the
electroweak scale $M_Z$ when the renormalization group equations (RGE) are
used to run the $\alpha_i$ up to $M_G$, We restrict our analysis here to the
simplest such model where R-parity is conserved and there are universal soft
breaking masses at $M_G$ ({\it{i.e.}}, mSUGRA). Such models depend on four parameters
and one sign : $m_0$ the universal soft breaking scalar mass at $M_G$;
$m_{1/2}$ the universal gaugino mass at $M_G$. $A_0$
the universal cubic soft breaking mass at $M_G$; $\tan\beta\equiv\,<H_2>/<H_1>$
where $<H_{1,2}>$ gives rise to ($d$,$u$) quark masses, and the sign of $\mu$, the
Higgs mixing parameter which appears in the $\mu H_1H_2$ contribution in the
superpotential. (Note that the gluino mass
scales approximately with $m_{1/2}$, {\it{i.e.}}, $m_{\tilde g}\stackrel{\sim}{=}2.4\,m_{1/2}$.) 
No assumptions are made on the nature of the GUT group which
breaks to the SM group at $M_G$. The model used here is the same as that used in LHC
analyses by ATLAS and CMS\cite{atlas_tdr,cms_susy}. Over almost all of the parameter space, the
lightest neutralino ($\tilde\chi^0_1$) is the lightest supersymmetric particle
(LSP), and is a natural candidate for cold dark matter\cite{lsp_cdm}.

In the present study,
we fix \azero\ = 0 and the sign of $\mu$ to be positive
(\mupos) for simplicity, and choose
 $\tan\beta$ = 3, 10, and 30.
Here the \ISAJET\ sign convention for $\mu$ \cite{isajet} is used.
The top quark mass is set to 175 \mgev.
We restrict the parameter space so that
the lighter third generation squarks ($\sbottomone$ and $\stopone$)
remain heavier than the lightest chargino $\schionepm$ and the next to lightest
neutralino $\schitwozero$,
and also  develop cuts
sensitive to gluinos and squarks.
These would decay to the SM particles plus the \lsp.
For example,
$\gluino \to  q \bar{q}^{\prime} \schionepm$,
$\squark_{L} \to q^{\prime} \schionepm$, 
followed by
$\schionepm \to q \bar{q}^{\prime} \lsp$
or  $\schionepm \to \ell^\pm \nu \lsp$.
The \lsp\ then would  pass through  the detector without interaction.
Thus, the experimental signatures of pair-produced
squarks and gluinos are multi jets and
appreciable missing energy associated with either no lepton or some leptons.
It should be noted that
the event selection with a large jet multiplicity presented later
in this paper is not efficient to detect
the production of $\squark_{R} \squarkb_{R}$, because
each right chiral scalar quark dominantly decays to a quark and a \lsp.

\section{Monte Carlo Simulation}
\label{sec:mc}
We use
\ISAJET\ \cite{isajet}  for SUSY and \ttb\ events and
\PYTHIA\ \cite{pythia} for
all other SM processes ($W/Z$ + jets, dibosons, QCD events)
along with \TAUOLA\  \cite{tauola}
and CTEQ3L parton distribution functions \cite{cteq3l}.
As for SUSY events, we generate all processes for
the analyses described in Section \ref{sec:results}. For detector simulation we
use \SHW\  \cite{shwsim},
 a simple detector simulation package developed for Run II SUSY/Higgs
workshop \cite{run2shw}. The particle identification and misidentification efficiencies
are parameterized to an expectation for Run II
based on the CDF/D\O\ measurements at Run I in 1992-96. 
The \SHW\ code provides the following objects:
electron ($e$) with isolation, muon ($\mu$) without isolation, 
hadronically decaying tau lepton ($\tau_h$), 
photon ($\gamma$), jets, and calorimeter-based \met.
\met\ is the energy imbalance in the directions
transverse to the beam direction
using the calorimeter energies in an event \cite{CDFcoordinate}. 
We modify the \SHW\ code to provide a muon with the isolation and
the \met\ correction due to muon(s).
The pseudorapidity ($\eta$) coverage 
is $|\eta|<2.0$ for $e$ and $\gamma$ \cite{run2shw}. 
For $\mu$, $\tau_h$ and tracks $|\eta|$ is $<1.5$, and 
for jets it is $<4.0$ \cite{run2shw}.
Jets are formed  with a cone size of 
$\Delta R \equiv \sqrt{\Delta \eta^2 + \Delta \phi^2}$ = 0.4.
A non-instrumented region of the detector is also simulated
as a geometical acceptance for each object.
(For example,
\SHW\ will reject a particular object at a rate of 10\%, if the
fiducial volume in a given pseudorapidity coverage is 90\%.) 
The isolation for an electron is defined to be
the calorimeter energy (excluding the electron energy)
within $\Delta R$ = 0.4
which is less than 2 GeV.
The isolation for a muon is defined as a scalar sum
of track momenta (excluding the muon momentum)  
within $\Delta R$ = 0.4 to be less than 2 GeV.
It should be noted that
hadronically decaying taus  ($\tau_h$) are treated as a jet.

Throughout the paper, the leptons and jets are selected with $p_{T}^{\ell}>15$
 GeV and $E_{T}^{j}>15~{\rm GeV}$, and  the reach in mass is obtained
as $5 \sigma$ in a significance ($\equiv N_S/\sqrt{N_B}$)
for 15 \invfb\ and 30 \invfb\ at the Tevatron and 30 \invfb\ at the Tripler.
Here $N_S$ ($N_B$) is the number of signal (background) events
after a set of selection cuts.

\section{Results}
\label{sec:results}
\begin{figure}
    \vspace{1.0cm}
    \begin{center}	
    \leavevmode
    \epsfysize=7.0cm
    \epsffile[75 160 575 630]{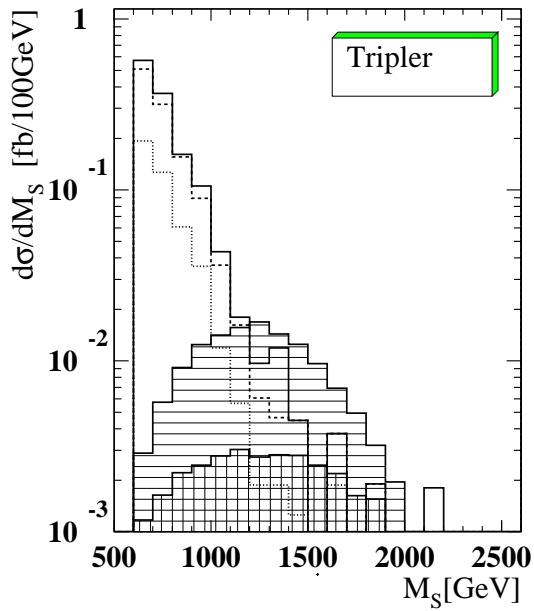} 
    \vspace{1.0cm}
    \caption{
	Distributions of $M_S$ for various SM processes
	and SUSY events with $M_S > 600$ GeV and $\mbox{${E\!\!\!\!/_T}$}
	 > 200\, {\rm GeV}$
	in the jets + \mbox{${E\!\!\!\!/_T}$} analysis at Tripler.
        Dotted, dashed and solid lines are
	cumulative contributions from \mbox{$t\overline{t}$},
	$W/Z$/dibosons,  
	and QCD processes, respectively.
	Horizontally and vertically hatched histograms are
	 for SUSY events ($\tan\beta = 3$)
	with $m_{\mbox{$\tilde{q}$}} \simeq m_{\mbox{$\tilde{g}$}} =$ 
	800 GeV and 1000 GeV,
	respectively.
	The final cut on $M_S$ is set at 1000 GeV.}
    \label{fig:trip_metjet_totet}
    \end{center}
\end{figure}

We consider first the jets + \mbox{${E\!\!\!\!/_T}$} channel and 
proceed to optimize the cuts for SUSY events where
 $m_{\mbox{$\tilde{q}$}} \simeq m_{\mbox{$\tilde{g}$}}$.
Our  optimized selection is 
(a)~$N_j \geq 6$;
(b)~veto on isolated leptons ($e$ or $\mu$); 
(c)~$\mbox{${E\!\!\!\!/_T}$} > 200$ GeV;
(d)~minimum azimuthal angle between the $\mbox{${E\!\!\!\!/_T}$}$ direction and any 
jet ~$\Delta \phi^{\rm min}> 30^{\circ}$; 
(e)~$M_S \equiv \mbox{${E\!\!\!\!/_T}$} + \sum_{jet} E_{T}^{j} > 1000\,{\rm GeV}$. 
Figure \ref{fig:trip_metjet_totet}
shows the  distributions in  $M_S$ 
for \mbox{$t\overline{t}$},  $W/Z$+jets, dibosons, and QCD  events.
The SUSY events
are also superimposed in the same figure.
We require in our analysis  $N_S\,\gtsim\,30$ events.
Using these cuts the total SM background is 7.0 fb (Table~\ref{background-table}).

\begin{figure}
    \begin{center}
    \leavevmode
    \epsfysize=7.0cm
    \epsffile[75 160 575 630]{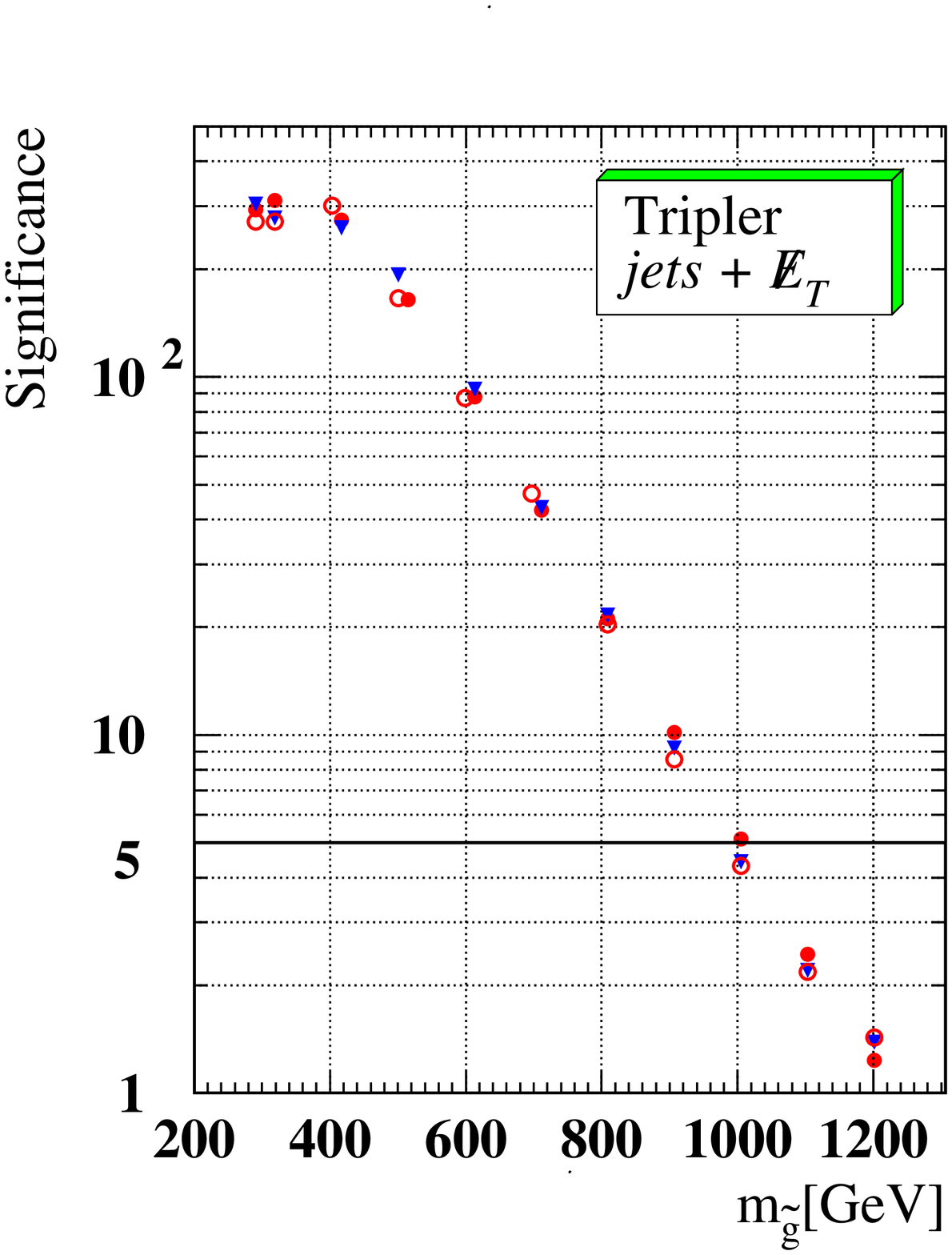} 
    \vspace{1.0cm}
    \caption{
	Significance as a function of $m_{\mbox{$\tilde{g}$}}$
	($m_{\mbox{$\tilde{q}$}} \simeq m_{\mbox{$\tilde{g}$}}$)
	for
	$\tan\beta$ = 3 (filled circles), 10 (down triangles), 
	and 30 (open circles) 
	in jets + \mbox{${E\!\!\!\!/_T}$}\ channel at the Tripler.}
    \label{fig:trip_metjet_tanB}
    \end{center}
\end{figure}

Significances for SUSY events ($m_{\mbox{$\tilde{q}$}} \simeq m_{\mbox{$\tilde{g}$}}$)
are plotted as 
 function of $m_{\mbox{$\tilde{g}$}}$
in Fig.~\ref{fig:trip_metjet_tanB}. We see that 
the  reach in the gluino mass is $\sim$ 1000 GeV
(corresponding to $m_{1/2} \simeq 420$ GeV). Also 
there is no significant dependence between the 
 $\tan\beta$ values of 3, 10 and 30, {\it{i.e.}}, the Tripler is sensitive to high
 $\tan\beta$.

\begin{figure}
    \begin{center}
    \leavevmode
    \epsfysize=7.0cm
    \epsffile[75 160 575 630]{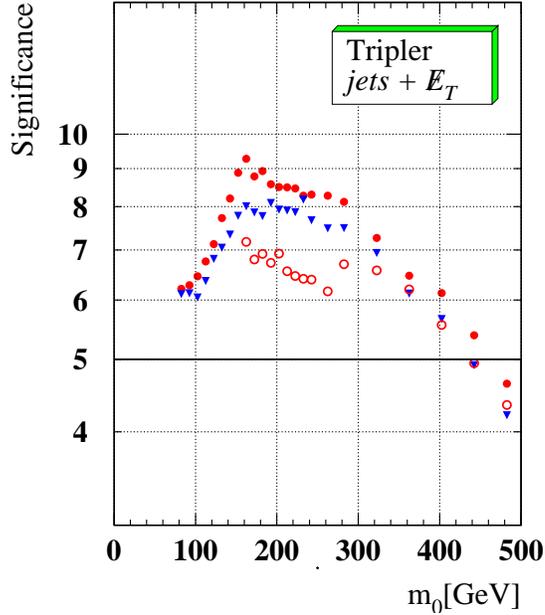} 
    \vspace{1.0cm}
    \caption{
	Significance  as a function of $m_0$ 
	 at $m_{1/2} = 410$ GeV ($m_{\tilde g}\simeq$ 980 GeV) for
	$\tan\beta$ = 3 (filled circles), 10 (down triangles), 
	and 30 (open circles) in jets + \mbox{${E\!\!\!\!/_T}$} channel at 
	the Tripler. $m_0\,\stackrel{<}{\sim}\,150$ GeV is theoretically forbidden for $\tan\beta$
	= 30.}
    \label{fig:trip_metjet_m0scan}
    \end{center}
\end{figure}

In
Fig.~\ref{fig:trip_metjet_m0scan} we plot the 
 significance
as a function of $m_0$ at $m_{1/2} = 410$ GeV. We see that at the highest
gluino mass, the Tripler is sensitive to relatively large $m_0$, {\it{i.e.}}, 
$m_0 \,\ltsim\, 450$ GeV for this channel.

In
Fig.~\ref{fig:trip_metjet_m0scan}, there are three distinct regions:
(i) $m_0 \,\ltsim \,150$ GeV,
(ii) $150\,\ltsim \,m_0\,\ltsim\,300$ GeV,
(iii) $m_0\,\gtsim\,300$ GeV.
For $m_0\,\gtsim$ 300 GeV, all sleptons are heavier than
$\tilde\chi^{\pm}_1$ and $\tilde\chi^{0}_2$. 
The jet multiplicity in the SUSY events is determined by
the $W$, $Z$, and light Higgs boson ($h$) decays in 
$\tilde\chi^{\pm}_1\rightarrow W \tilde\chi^0_1$ and $\tilde\chi^{0}_2 \rightarrow h \tilde\chi^0_1
 / Z \tilde\chi^0_1$,
whose  branching ratios are independent of $m_0$.
Thus there is no change in the event topology, but
the cross section for $\mbox{$\tilde{q}$}\mbox{$\bar{\tilde{q}}$}$ and 
$\mbox{$\tilde{g}$}\mbox{$\tilde{q}$}$ decrease
as  $m_0$ ({\it{i.e.}}, squark masses) increases.
For  $150\,\ltsim\ m_0\,\ltsim\ 300$ GeV,
the $m_0$ dependence becomes somewhat gradual. 
This is because, as  $\tilde\tau_1$ (and $\tilde e_{R}$) 
gets lighter than $\tilde\chi^{\pm}_1$,
 the decay mode $\chi^{\pm}_1\rightarrow\tilde\tau_1 \nu$
starts competing with  $\tilde\chi^{\pm}_1 \rightarrow W^{\pm}\tilde\chi^0_1$.
Thus
the jet multiplicity in the SUSY events invloving $\tilde\chi^{\pm}_1$ decay
mode
is reduced to affect  its event acceptance (with $N_j \geq 6$).
In contrast, $\tilde\chi^{0}_2 \rightarrow \tau \tilde\tau_1$ decay
(competing with $\tilde\chi^0_2\rightarrow h \tilde\chi^0_1 / Z \tilde\chi^0_1$ especially for
$\tan\beta$ = 10 and 30)
does not alter the jet multiplicity.
We notice the significance  has
a $\tan\beta$ dependence at a fixed $m_0$.    
This can be explained by
(a)~a change of the third-generation squark masses
(especially the $\tilde t_1$), resulting
	in the change of squark production cross sections  and
(b)~an enhancement of branching ratio
for $\tilde\chi^{\pm}_1 \to \tilde \tau_1 \nu$ in
large $\tan\beta$ region, The decay mode
 $\tilde\chi^{\pm}_1\rightarrow W^{\pm} \tilde\chi^0_1$ is dominant in low  $\tan\beta$ region,
resulting in the change of jet multiplicity.
A  characteristic change for $m_0\,\ltsim\ 150$ GeV 
at $\tan\beta$ = 3 and 10, where
$\tilde e_{L}$ and $\tilde \nu$ become lighter than $\tilde\chi^{\pm}_1$ and
$\tilde\chi^0_2$),
is explained by
a monotonic decrease  of rates of $\tilde\chi^{\pm}_1\rightarrow
W^{\pm}\tilde\chi^0_1$
and $\tilde\chi^0_2\rightarrow h \tilde\chi^0_1$ decays as $m_0$ decreases.
Thus the significance of the SUSY events
with $N_j \geq 6$ is degraded.

\begin{figure}
    \begin{center}
    \leavevmode
    \epsfysize=7.0cm
    \epsffile[75 160 575 630]{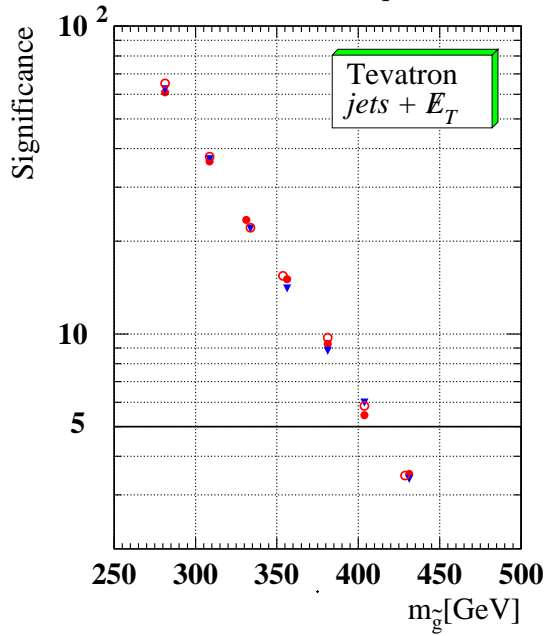}
    \vspace{1.0cm}
    \caption{
	Significance for 15 fb$^{-1}$ of luminosity as a function of $m_{\gluino}$ 
	($m_{\squark} \simeq m_{\gluino}$) for
	$\tan\beta = 3$ (filled circles), 10 (down triangles), and
	30 (open circles) in jets + \met\ channel at the Tevatron.}
    \label{fig:tev_metjet_tanB3_10_30}
    \end{center}
\end{figure}

Figure~\ref{fig:tev_metjet_tanB3_10_30} gives a comparison of what might be
expected at Run II at the Tevatron with 15 fb$^{-1}$. Our selection cuts in this case were:
(a)~$N_j \geq 4$;
(b)~veto on isolated leptons;
(c)~$\met > 100\ \gev$;
(d)~$\Delta \phi^{\rm min} > 30^{\circ}$;
(e)~$M_{S_2}(\equiv \met\ + E_T^{j_1} + E_T^{j_2}) > 350\ \gev$\cite{tev2000_mrenna}.
The total SM background is 73 fb (Table~\ref{background-table}).

One sees that the maximum
reach for the jets +\mbox{${E\!\!\!\!/_T}$} channel in 
Fig.~\ref{fig:tev_metjet_tanB3_10_30} is 410 GeV in gluino mass, which rises
to 440(460) GeV for 15(30) fb$^{-1}$ of data when $m_{\tilde q}<m_{\tilde g}$. 
(These results are consistent with previous
Tevatron studies\cite{tev2000_baer,Baer_large_tanB}.) The latest bound on the 
$\tilde\chi^{\pm}_1$ mass of $\gtsim\ 103$ GeV from LEPII\cite{lep} 
requires $m_{\tilde g}\,\gtsim\ 420$ GeV,
 since we have 
$m_{\tilde \chi^{\pm}_1}\simeq m_{\tilde g}/3$ from gaugino unification. In
addition, one may show that Run II will also be able to sample limited range of $m_0$, {\it{i.e.}}, for 
$m_0\,\ltsim$ 200 GeV for $m_{\tilde g}=$ 420 GeV.  Thus there is a significant
improvement in going from the Tevatron to the Tripler.

We consider next the $1\ell$ + jets + \met\  channel. This channel gives the largest reach for the LHC,
and we will see that there are regions of SUSY parameter space where the
discovery reach for gluinos is also improved. Here we select events with
(a)~$N_j \geq 4$;
(b)~$N_{\ell} = 1$;
(c)~$\met > 200\ \gev$;
(d)~$\Delta \phi^{\rm min} > 30^{\circ}$;
(e)~$M_T(\equiv \sqrt{ 2 \met p_T^{\ell} 
[1 - \cos\Delta\phi(\ell,\met) ]}) > 160\ \mgev$;
(f)~$M_{S}   > 600\ \gev$.
The $M_T$ cut is applied to remove $W$ events.
The SM background is 0.32 fb (Table~\ref{background-table}).

\begin{figure}
    \begin{center}
    \leavevmode
    \epsfysize=7.0cm
    \epsffile[75 160 575 630]{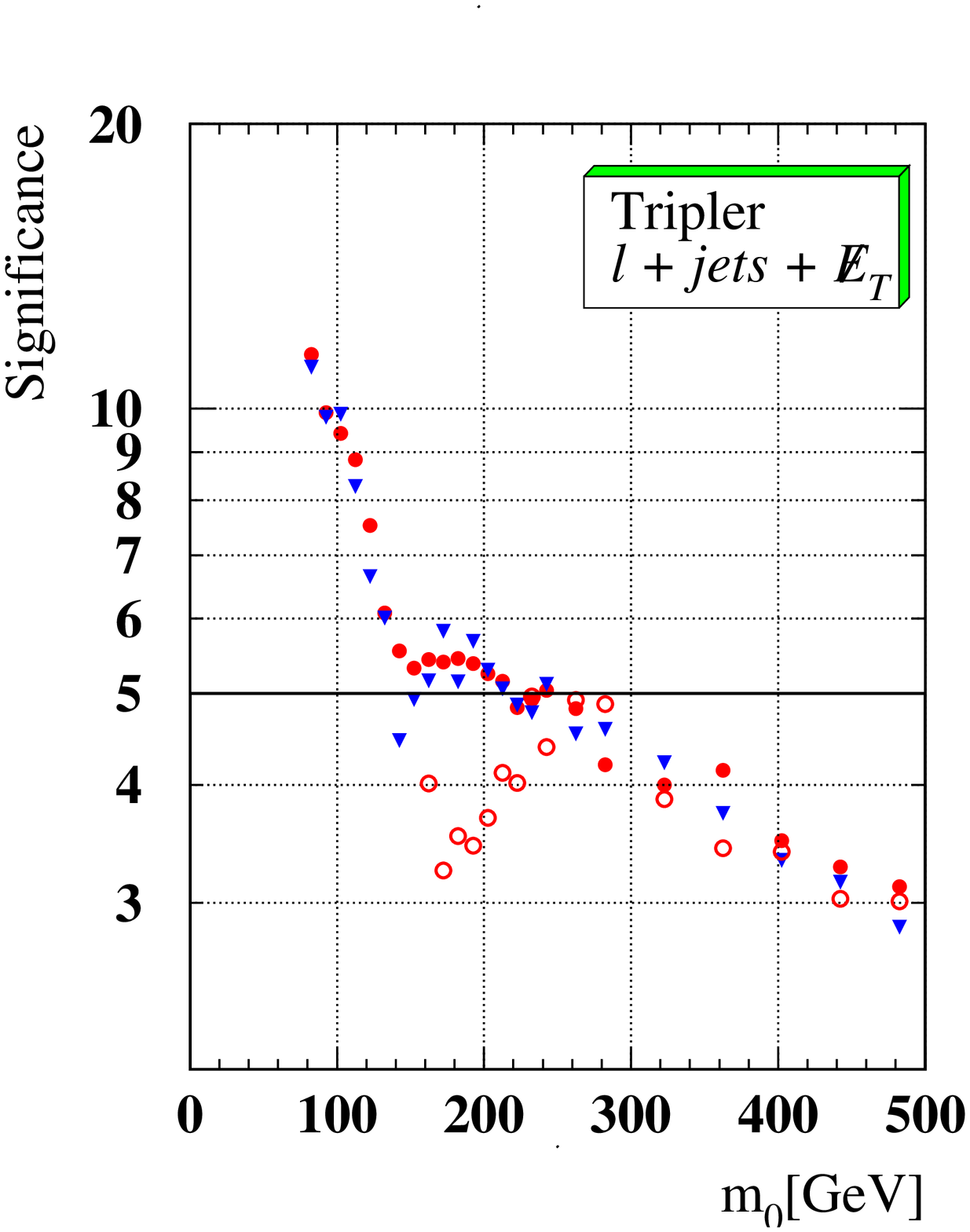}
    \vspace{1.0cm}
    \caption{
	Same as in Fig. \ref{fig:trip_metjet_m0scan}
	($\mhalf\ = 410\ \mgev$), but
	in  $1\ell$ + jets + \met\ channel.}
    \label{fig:trip_1lmetjet_m0scan}
    \end{center}
\end{figure}

In Fig. \ref{fig:trip_1lmetjet_m0scan}, we compare the $\mzero$ dependence
of the significance for $\tan\beta$ = 3, 10 and 30 at $\mhalf = 410\ \mgev$
($m_{\tilde g}\simeq 980$ GeV). We see here the $m_0$ reach is not as large as in the 
jets + \met\  channel.
In this parameter space, the  
$\slepton_{L}$ and the $\snu$  are  lighter than
the $\schionepm$ and the $\schitwozero$ when  $\mzero\ \ltsim\ 150\ \mgev$.
Thus, the branching ratios of
$\schionepm \to \ell \snu\ (\slepton_{L} \nu)$ and
$\schitwozero \to \ell \slepton$ increase as $\mzero$ decreases,
and
the significance in the $1\ell$ + jets + \met\  channel 
is  improved dramatically and is significantly higher than for the jets + \met\ 
channel (See Fig. \ref{fig:trip_metjet_m0scan}).
An interesting feature occurs at  $\mzero\ \simeq\ 140\ \mgev$
which distinguishes between the $\tan\beta$ = 3 and 10 scenarios.
The SUSY particle masses for these two $\tan\beta$ values are very close, except
for the $\stauone$ mass.
The $\stauone$ mass at $\tan\beta = 10$ is lighter,  so that
the branching ratios of $\schionepm \to \stauone \nu$ and
$\schitwozero \to \tau \stauone$ are larger
to decrease the $1\ell$ + jets + \met\ signature.

\begin{figure}
    \begin{center}
    \leavevmode
    \epsfysize=7.0cm
    \epsffile[75 160 575 630]{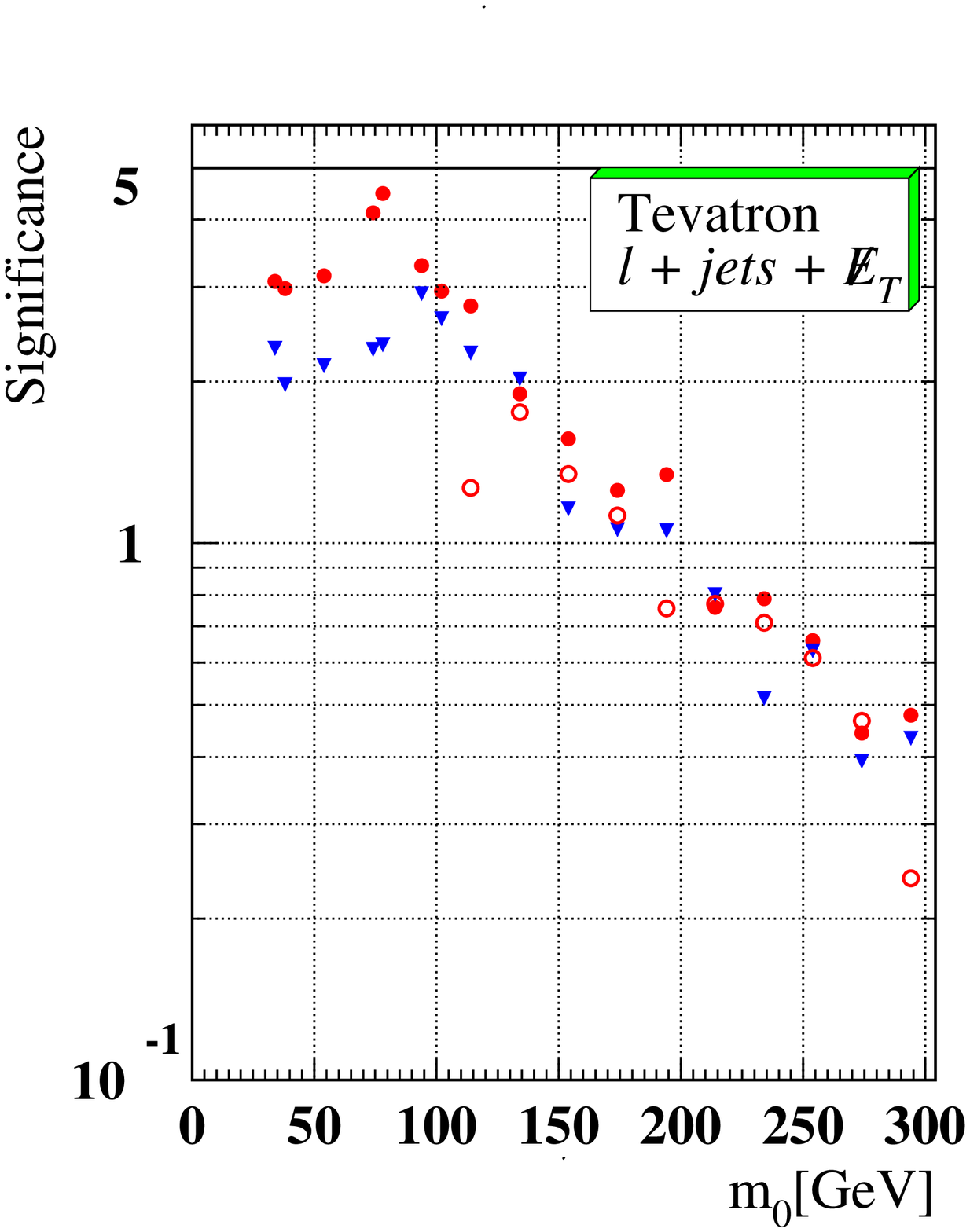}
    \vspace{1.0cm}
    \caption{Significance for 15 fb$^{-1}$ of luminosity as a function of $m_0$  for 
$\tan\beta$ = 3 (filled circles), 10 (down triangles), 
and 30 (open circles) 
in $1\ell$ + jets + \met\ channel at the Tevatron for 
	  $m_{1/2} = 160\ \mgev$ ($m_{\tilde g}\simeq$ 420 GeV).}
    \label{fig:tev_1lmetjet_m0scan}
    \end{center}
\end{figure}

As before, for comparison, we show the significance for $m_{1/2} = $ 160 GeV 
in this channel for the Tevatron in 
Fig. \ref{fig:tev_1lmetjet_m0scan}.  
The event selections for this figure was made with the following cuts:
(a)~$N_j \geq 2$;
(b)~$N_{\ell}$ = 1;
(c)~$\met > 40\ \gev$;
(d)~$\Delta \phi^{\rm min} > 30^{\circ}$;
(e)~$M_T < 50\ \mgev$ or $> 110\ \mgev$;
(f)~$M_{S_2} > 350\ \gev$.
The $M_T$ cut is applied to remove $W$ events.
The SM background size is 70 fb (Table~\ref{background-table}). The significance is found to be always below 5$\sigma$ for 
the entire region of parameter space. 

\begin{figure}
    \begin{center}
    \leavevmode
    \epsfysize=7.0cm
    \epsffile[75 160 575 630]{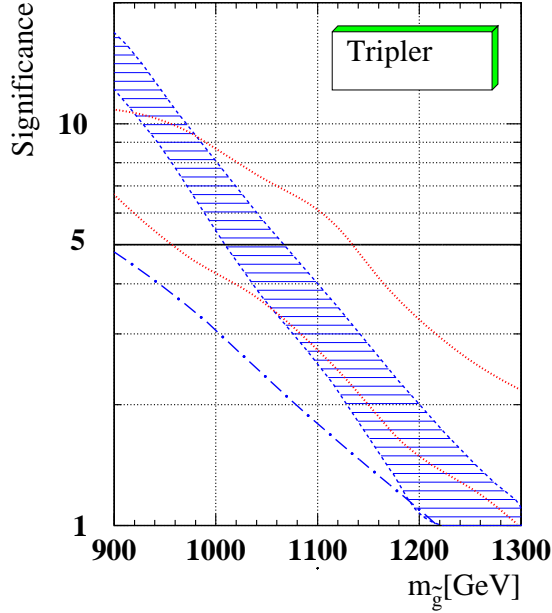} 
    \vspace{1.0cm}
    \caption{
	Significance as a function of $m_{\gluino}$ for
	$\tan\beta$ = 3
	in jets + \met\  (hatched region bounded by the dashed lines) and 
	$1\ell$ + jets + \met\ (region bounded by the dotted lines) channels at the Tripler.
	Ranges scanned are $360 \leq \mhalf\ \leq 540\ \mgev$ and
	$100 \leq \mzero\ \leq 260\ \mgev$. The dot-dashed line represents
	jets + \met\ channel
for $m_0$= 650 GeV and $\tan\beta = 3$.}
    \label{fig:trip_metjet_1lmetjet}
    \end{center}
\end{figure}

From Figs. \ref{fig:trip_metjet_m0scan} and \ref{fig:trip_1lmetjet_m0scan},
the significance  in jets + \met\ and $1\ell$ + jets + \met\ channels
appear to be maximized
at $\mzero$ = 140-160 \mgev\ and 
100 GeV  respectively,
for $\mhalf = 410\ \mgev$.
To obtain the strongest reaches we therefore 
systematically scan mSUGRA points at  $\tan\beta = 3$ for 
$\mhalf = \{ 360, 400, 440, 470, 500, 540 \}$ GeV and
$\mzero = \{ 100, 140, 180, 220, 260 \}$ GeV.
Figure  \ref{fig:trip_metjet_1lmetjet} shows significance 
as a function of the gluino mass in both
jets + \met\ (hatched region bounded by the dashed lines) and 
$1\ell$ + jets  + \met\ (region bounded by the dotted lines)  analyses for the above mSUGRA points.
We see that the strongest reach  
in the jets~+~\met\  channel is $m_{\tilde g}\simeq$ 1060 GeV
($\mhalf\ \simeq 440\ \mgev$) and  $m_{\tilde g}\simeq$ 1140 GeV 
($\mhalf\ \simeq 480\ \mgev$) in  the $1\ell$ + jets + \met\ channel. 
The dot-dashed line represents jets + \met\ channel
for $m_0$= 650 GeV and $\tan\beta = 3$. Even for this large  $m_0$, we can see that the 
5$\sigma$ significance can be achieved  for $m_{\tilde g}\simeq$ 900 GeV 
($m_{\tilde q_{1,2}}\simeq$ 970 GeV, where $m_{\tilde q_{1,2}}$ are the squark
masses of the first two generations).

\section{Conclusion} We have studied the signals for gluinos and squarks
within the framework of mSUGRA models in the jets + \met\ and 
$1\ell$ + jets + \met\ channels for the Tripler $p{\bar p}$ accelerator with
$\sqrt s=5.4$ TeV. The Tripler would have a maximum reach of 
$m_{\tilde g}\simeq$ 1140 GeV 
with 30 fb$^{-1}$ in the $1\ell$ + jets + \met\ channel (for $m_0\simeq 100$ GeV,
$\tan\beta$=3-10) and $m_{\tilde g}\simeq$ 1060 
GeV in the jets~+~\met\ channel (for
140 $\ltsim\ m_0\ltsim$ 200 GeV, $\tan\beta=3$). This
gluino mass reach is comparable to the Tripler's reach of 380 GeV 
 chargino (with
40 fb$^{-1}$) in the trilepton channel\cite{tripler_wisc} via direct
chargino-neutralino ($\tilde\chi^{\pm}_1$-$\tilde\chi^{0}_2$) production, since gaugino unification implies
$m_{\tilde \chi^{\pm}_1}\simeq m_{\tilde g}/3$.
The above results can
be compared with 440(460) GeV  for the jets + \met\ channel 
for 15(30) fb$^{-1}$ of luminosity at the Tevatron. For  $m_{\tilde g}\simeq$
980 GeV, the Tripler covers relatively large values of
$m_0$, {\it{i.e.}}, to $m_0\ltsim\ $420 GeV in the jets + \met\ channel. Note
also, from Figs. \ref{fig:trip_metjet_tanB} and \ref{fig:trip_metjet_m0scan}, that this  gluino and $m_0$ reach of the Tripler is valid for large tan$\beta$
while the trilepton analysis \cite{tripler_wisc} is sensitive only for small
tan$\beta$ ($e.g.$,
tan$\beta$=3).

In the above analysis we have set $A_0$=0. The results for the maximum Tripler
reach are not very sensitive to $A_0$. Thus there is almost no change for
$A_0>0$ and for $A_0=-1000~{\rm GeV}$, tan$\beta$=3, the gluino reach is increased by
about 20 GeV in the $1\ell$ + jets + \met\ channel.

In SUGRA models of this type, the $\tilde\chi^0_1$ is the LSP and hence is the
main candidate for cold dark matter. The astronomical constraints on the amount
of relic neutralinos generally implies $m_0\,\ltsim\,$200 GeV, for 
$m_{1/2}\ltsim$ 350-400 GeV. For higher $m_{1/2}$,
coannihilation effects dominate \cite{efo} and for high $\tan\beta$, $m_0$ can rise to 
$\ltsim$ 400-500 GeV \cite{ads}. Thus the Tripler would be sensitive to
much of the cosmologically interesting part of the parameter space.

The LHC gluino reach is $m_{\tilde g}\,\ltsim$ 2.5 TeV\cite{atlas_tdr,cms_susy}, which
is much higher than the Tripler. The Tripler is however complementary to the LHC
in that the production of squarks and gluinos go in part through different
channels, as are the detector signals for the charginos and neutralinos. Thus
provided SUSY lies sufficiently low to be seen at the Tripler, the two
accelerators would be sensitive to different supersymmetric interactions.
\section*{Acknowledgements}

We thank Muge Karagoz for her participation in the 
earlier stage of our analysis.
This work was supported in part by
DOE grant Nos. DE-FG03-95ER40917, DE-FG03-95ER40924 and
NSF grant No. PHY-9722090.

%\begin{table}
%\caption{Pseudorapidity coverages for objects in the \SHW\ simulation
%package.}\label{tab:shw_eta}
%\begin{center}
%\begin{tabular}{ll}
%\hline\hline
%Object		& $| \eta |$ Coverage \\
%\hline
%electrons ($e$), photons ($\gamma$) & $< 2.0$ \\
%muons ($\mu$)		& $< 1.5$ \\
%taus ($\tau_h$)		& $< 1.5$	\\
%tracks		& $< 1.5$	\\
%jets ($j$)		& $< 4.0$	\\
%\hline\hline
%\end{tabular}
%\end{center}
%\end{table}

\begin{table}\caption{The background cross sections in fb
(after cut) of  
jets + \met\ and $1\ell$ + jets + \met\ channels.  
Cuts are specified in the text.}\label{background-table}
\begin{center}
\begin{tabular}{|l|c|c|c|c|}
\hline\hline
&\multicolumn{2}{c|}{Tripler} &\multicolumn{2}{c|}{ Tevatron} \\\hline
%&\multicolumn{2}{c|}{$\sigma$ (in fb) after cut}&
%\multicolumn{2}{c|}{$\sigma$ (in fb) after cut}\\\hline
Process&jets + \met\ &$1\ell$ + jets + \met\ &jets + \met\ &$1\ell$ + jets + \met\ \\\hline
$t{\bar t}$&2.5&0.06&25&14\\
$W$+jets&1.1&0.21&23&48\\
$Z$+jets&1.2&0.05&15&6\\
diboson&0.0&0.0&1&2\\
QCD&2.2&0.0&9&0.0\\\hline Total&7.0&0.32&73&70\\ \hline\hline
\end{tabular}
\end{center}\end{table}


\begin{thebibliography}{99}

\bibitem{tripler}
 	 P. McIntyre, E. Accomando, R. Arnowitt, B. Dutta, T. Kamon, 
	and A. Sattarov, 
	hep-ex/9908052 (1999).

\bibitem{BNLmagnet}
        R. Gupta,
        ``Common Coil Magnet System for VLHC,''
        Proc. of 1999 Part. Acc. Conf.,
        New York, p.3239 (1999).

\bibitem{FNALmagnet}
        G. Ambrosio \etal,
        ``Conceptual Design of the Fermilab Nb$_{3}$Sn High Field Dipole
        Model,''Proc. of 1999 Part. Acc. Conf.,
        New York, p.~174 (1999).

\bibitem{LBNLmagnet}
        K. Chow \etal,
        ``Fabrication and Test Results of
        a Nb$_{3}$Sn Superconducting Racetrack Dipole Magnet,''Proc. of 1999 Part. Acc. Conf.,
        New York, p.~171 (1999).

\bibitem{TAMUmagnet}
        C. Battle \etal,
        ``Optimization of Block-coil Dipoles for
        Hadron Colliders,''Proc. of 1999 Part. Acc. Conf.,
        New York, p.~2936 (1999).


\bibitem{bag} American Linear Collider Working Group (J. Bagger \etal),
``The Case for a 500-GeV $e^{+}e^{-}$ Linear Collider,'' hep-ex/0007022
(2000). 

\bibitem{tripler_wisc} 
 	 V. Barger, K. Cheung, T. Han, C. Kao, T. Plehn, R-J. Zhang, 
%		``Physics Potential of a Tevatron Tripler,''
	 \Journal{\PLB}{478}{224}{2000}.


\bibitem{atlas_tdr}
	ATLAS: Detector and Physics Performance Technical Design Report, 
	vol. 1,
	CERN-LHCC-99-14, ATLAS-TDR-14 (1999); vol. 2, CERN-LHCC-99-15,
	ATLAS-TDR-15 (1999).

\bibitem{cms_higgs} K. Lassila-Perini,``Discovery Potential of the  Standard Model Higgs
in CMS at the LHC,''  CERN-THESIS-99-3(1998).

\bibitem{smhiggs_trivbound}
        U.M. Heller, M. Klomfass, H. Neuberger, and P. Vranas,
%               ``Numerical Analysis of the Higgs Mass Triviality Bound,''
        \Journal{\NPB}{405}{555}{1993}.
	
\bibitem{mssm} For reviews on SUSY and the MSSM, see $e.g.$,
        H.P. Nilles, \Journal{\PR}{110}{1}{1984} and
        H.E. Haber and G.L. Kane, \Journal{\PR}{117}{75}{1995}.

\bibitem{msugra}  
        A.H. Chamseddine, R. Arnowitt, and P. Nath, 
%		``Locally Supersymmetric Grand Unification,"  
		\Journal{\PRL}{49}{970}{1982}; 
        R.~Barbieri, S. Ferrara, and C.A. Savoy, 
%		``Gauge Models with Spontaneously Broken Local Supersymmetry,''
		 \Journal{\PLB}{119}{343}{1982}; 
        L. Hall, J. Lykken, and S. Weinberg, 
%		``Supergravity as the Messenger of Supersymmetry Breaking,'' 
		\Journal{\PRD}{27}{2359}{1983}. 
        For a review, see R. Arnowitt and P. Nath, \Journal{\NPB}{227}{121}{1983}. 


\bibitem{tev2000_baer}
        H. Baer, C.-H. Chen, C. Kao, and X. Tata,
%       ``Supersymmetry Reach of an Upgraded Fermilab Tevatron Collider,''
%       (W1Z2, W1W1 cases)
        \Journal{\PRD}{52}{1565}{1995};
        H. Baer, C. Chen, F. Paige, and X. Tata,
%       ``Supersymmetry Reach of Fermilab Tevatron Upgrades:
%       A Comparative Study,''
        \Journal{$ibid.$}{54}{5866}{1996}.

\bibitem{tev2000_mrenna}
        S. Mrenna, G.L. Kane, G.D. Kribs, and J.D. Wells,
        \Journal{\PRD}{53}{1168}{1996}.

\bibitem{tev2000}
        TeV2000 Group,
        ``Future ElectroWeak Physics at the Fermilab Tevatron:
        Report of the TeV2000 Study Group,''
        $eds.$ D. Amidei and R. Brock,
        Fermilab-Pub-96/082 (1996).

\bibitem{Baer_large_tanB}
        H.~Baer, C.-H. Chen, M. Drees, F. Paige, and X. Tata,
        \Journal{\PRL}{79}{986}{1997};
        \Journal{$ibid.$}{80}{642}{1998} (E);
%        ``Supersymmetry Reach of Tevatron Upgrades:
%        The Large $\tan\beta$ Case,''
        \Journal{\PRD}{58}{075008}{1998}.

\bibitem{Barger_large_tanB}
        V.~Barger, C. Kao, and T-J. Li,
%        ``Trilepton Signal of Minimal Supergravity
%        at the Tevatron Including $\tau$-lepton Contributions,''
%       hep-ph/9804451,
        \Journal{\PLB}{433}{328}{1998};
        V.~Barger and C. Kao,
%       ``Trilepton Signature of Minimal Supergravity at the
%       Upgraded Fermilab Tevatron,''
%
        \Journal{\PRD}{60}{115015}{1999}. % and hep/ph/9811489.

\bibitem{baer2000}H.~Baer,  M. Drees, F. Paige, P. Quintana, and X. Tata,
\Journal{\PRD}{61}{095007}{2000}.

\bibitem{run2shw_tau}
        K.T. Matchev and D.M. Pierce,
        \Journal{\PRD}{60}{075004}{1999};
%               Fermilab-Pub-99/078-T and hep-ph/9904282 (Phys. Rev. D);
        \Journal{\PLB}{467}{225}{1999};
%               Fermilab-Pub-99/209-T and hep-ph/9907505 (Phys. Lett. B);
        J.D. Lykken and K.T. Matchev,
        \Journal{\PRD}{61}{015001}{2000}.
%               Fermilab-Pub-99/034-T and hep-ph/9903238 (Phys. Rev. D.).

\bibitem{Accomando}
        E. Accomando, R. Arnowitt, and B. Dutta,
%       ``Trilepton Signal of Grand Unified Models at the Tevatron,''
%       (Non-universarity of m0)
%       hep-ph/9811300,
         \Journal{\PLB}{475}{176}{2000}.


\bibitem{anderson}
        G. Anderson, H. Baer, C.-H. Chen, and X. Tata,
%       ``The Search of Fermilab Tevatron Upgrades for
%       SU(5) Supergravity Models with Non-Universal Gaugino Masses,''
%       (Non-universarity of gauginos)
%       hepr-ph/9903370,
        \Journal{\PRD}{61}{095005}{2000}.

\bibitem{regina}
        R. Demina, J.D. Lykken, K.T. Matchev, and A. Nomerotski,
        \Journal{\PRD}{62}{035011}{2000}.

\bibitem{run2shw}
        S. Abel \etal,
        ``Report of the SUGRA Working Group For Run II of the Tevatron,''
        hep-ph/0003154 (2000).

%\bibitem{atlas_tdr}
%	ATLAS: Detector and Physics Performance Technical Design Report, 
%	vol. 1,
%	CERN-LHCC-99-14, ATLAS-TDR-14 (1999); vol. 2, CERN-LHCC-99-15,
%	ATLAS-TDR-15 (1999).
%
\bibitem{cms_susy}
        CMS Collaboration, S. Abdullin \etal,
        ``Discovery Potential for Supersymmetry in CMS,''
        hep-ph/9806366 (1998).

%\bibitem{msugra}  
%        A.H. Chamseddine, R. Arnowitt, and P. Nath, 
%		``Locally Supersymmetric Grand Unification,"  
%		\Journal{\PRL}{49}{970}{1982}; 
%        R. Barbieri, S. Ferrara, and C.A. Savoy, 
%		``Gauge Models with Spontaneously Broken Local Supersymmetry,''
%		 \Journal{\PLB}{119}{343}{1982}; 
%        L. Hall, J. Lykken, and S. Weinberg, 
%		``Supergravity as the Messenger of Supersymmetry Breaking,'' 
%		\Journal{\PRD}{27}{2359}{1983}. 
%        For a review, see R. Arnowitt and P. Nath, \Journal{\NPB}{227}{121}{1983}. 

\bibitem{lsp_cdm}
        H. Goldberg,
        \Journal{\PRL}{50}{1419}{1983};
        J. Ellis, J.S. Hagelin, D.V. Nanopoulos, K. Olive, and
        M. Srednicki,
        \Journal{\NPB}{238}{453}{1984}.
	
\bibitem{isajet} H. Baer, F. Paige, S.D. Protopopescu, and X. Tata,
        ``ISAJET 7.48: A Monte Carlo Event Generators for
        $pp$, $\bar{p}p$, and $e^+e^-$ Reactions,'' hep-ph/0001086 (2000).


\bibitem{pythia}
	T. Sj$\ddot{\rm o}$strand, 
	Commput. Phys. Commun. {\bf 82}, 74 (1994); 
	T. Sjostrand, P. Eden, C. Friberg, L. Lonnblad,
G. Miu, and S. Mrenna, hep-ph/0010017 (2000).

\bibitem{tauola}
        S. Jadach, J.H. K$\ddot{\rm u}$hn, and Z. W\c{a}s,
        \Journal{\CPC}{64}{275}{1991};
%       Commput. Phys. Commun. {\bf 64}, 275 (1991);
        M. Je$\dot{\rm z}$abek, Z. W\c{a}s, S. Jadach,
                and J.H. K$\ddot{\rm u}$hn,
        {\it ibid.} {\bf 70}, 69 (1992);
        S. Jadach, Z. W\c{a}s, R. Decker, and J.H. K$\ddot{\rm u}$hn,
        {\it ibid.} {\bf 76}, 361 (1993).
        We use version 2.5.

\bibitem{cteq3l}
 	CTEQ Collaboration, H. L. Lai \etal, 
%	``Global QCD Analysis and the CTEQ 
%	Parton Distributions,'' 
        \Journal{\PRD}{51}{4763}{1995}.
%
%	MSUHEP-41024 and hep-ph/9410404


\bibitem{shwsim}
	J. Conway,
	``User's Guide to the SHW Package,''
	http://www.physics.rutgers.edu/ $\sim$jconway/soft/shw/shw.html.
	We use version 2.3 with
	StdHep (version 4.06 - 
	http://www-pat.fnal.gov/stdhep.html) and CERNLIB 2000.


\bibitem{CDFcoordinate}
        In the CDF/D\O\ coordinate system, $\theta$ and $\phi$ are 
        the polar and azimuthal angles 
        with respect to the proton beam direction.

\bibitem{lep}I. Trigger, OPAL Collaboration, talk presented at the DPF 2000,
Columbus, OH; T. Alderweireld, DELPHI Collaboration, talk presented at the DPF 2000,
Columbus, OH.

\bibitem{efo}J. Ellis, A. Ferstl, and K. A. Olive, \Journal{\PLB}{481}{304}{2000}.

\bibitem{ads}R. Arnowitt, B. Dutta, and Y. Santoso, hep-ph/0010244 (2000). 
\end{thebibliography}
\end{document}